\documentclass[10pt,prx,aps,twocolumn,superscriptaddress]{revtex4-2}

\usepackage[usenames, dvipsnames]{xcolor}
\usepackage[utf8]{inputenc}
\usepackage{hyperref}
\hypersetup{colorlinks,allcolors=blue}
\usepackage{amsmath,revsymb,amssymb,mathtools}
\usepackage{graphicx}
\usepackage{xcolor}
\usepackage{bbold,comment}
\usepackage{braket,ulem}
\usepackage{booktabs}
\usepackage{float}
\usepackage{physics}
\usepackage{parskip}
\usepackage{multirow}
\usepackage{tabularx}
\usepackage{array}
\usepackage{pgf}
\usepackage{csquotes}
\usepackage{comment}
\usepackage{booktabs}
\usepackage{tabularx}

\newcolumntype{C}[1]{>{\centering\let\newline\\\arraybackslash\hspace{0pt}}m{#1}}
\newcolumntype{Y}{>{\centering\arraybackslash}X}

\begin{document}

\title{GHz-rate all-fiber active polarization state analyzer for quantum protocols}

\author{Andrea Pompermaier}
\email{andrea.pompermaier@phd.unipd.it}
\affiliation{Dipartimento di Ingegneria dell'Informazione, Universit\`a degli Studi di Padova, via Gradenigo 6B, IT-35131 Padova, Italy}

\author{Kannan Vijayadharan}
\affiliation{Dipartimento di Ingegneria dell'Informazione, Universit\`a degli Studi di Padova, via Gradenigo 6B, IT-35131 Padova, Italy}

\author{Costantino Agnesi}
\affiliation{Dipartimento di Ingegneria dell'Informazione, Universit\`a degli Studi di Padova, via Gradenigo 6B, IT-35131 Padova, Italy}
\affiliation{Padua Quantum Technologies Research Center, Universit\`a degli Studi di Padova, via Gradenigo 6B, IT-35131 Padova, Italy}

\author{Marco Avesani}
\affiliation{Dipartimento di Ingegneria dell'Informazione, Universit\`a degli Studi di Padova, via Gradenigo 6B, IT-35131 Padova, Italy}
\affiliation{Padua Quantum Technologies Research Center, Universit\`a degli Studi di Padova, via Gradenigo 6B, IT-35131 Padova, Italy}

\author{Giuseppe Vallone}
\affiliation{Dipartimento di Ingegneria dell'Informazione, Universit\`a degli Studi di Padova, via Gradenigo 6B, IT-35131 Padova, Italy}
\affiliation{Padua Quantum Technologies Research Center, Universit\`a degli Studi di Padova, via Gradenigo 6B, IT-35131 Padova, Italy}

\author{Paolo Villoresi}
\affiliation{Dipartimento di Ingegneria dell'Informazione, Universit\`a degli Studi di Padova, via Gradenigo 6B, IT-35131 Padova, Italy}
\affiliation{Padua Quantum Technologies Research Center, Universit\`a degli Studi di Padova, via Gradenigo 6B, IT-35131 Padova, Italy}

\begin{abstract}
    
Active selection of the measurement basis underpins quantum protocols including device-independent quantum key distribution, quantum teleportation with active feed-forward, and Bell tests. 
High-speed operation is crucial to minimize the latency between consecutive measurement choices, enabling faster protocol execution and higher achievable communication rates.
Here, we demonstrate a GHz-rate all-fiber polarization state analyzer enabling active, trial-by-trial reconfiguration of the measurement basis, which we validate by performing a CHSH Bell test. 
The state analyzer is based on a fiber Sagnac interferometer incorporating a lithium niobate electro-optic phase modulator, built entirely using off-the-shelf fiber-optic components. Operating at a nominal repetition rate of 1 GHz, the system performs dynamic polarization measurements for the CHSH Bell test, achieving polarization visibilities up to 99\% and a violation of $S=2.6975 \pm 0.0005$, certifying entanglement at an unprecedented rate. Furthermore, the system demonstrates excellent long-term stability, preserving the Bell-inequality violation for more than 6 hours without realignment. These results establish the proposed state analyzer as a versatile, scalable platform for quantum communication protocols requiring fast, reconfigurable polarization-state measurements.

\end{abstract}

\maketitle

\section{Introduction}

In the domain of quantum communication, photons are the preferred carriers of quantum information, encoding the qubit in a particular degree of freedom of the photon, such as polarization \cite{Kwiat1995Polarization}, path \cite{Politi2008Path}, time-bin \cite{Brendel1999TimeBin}, frequency-bin \cite{Olislager2010Frequency} and orbital angular momentum \cite{Mair2001OAM}. Among these, polarization encoding stands out for its experimental simplicity and the maturity of commercial optical components, offering a robust solution for both fiber and free-space quantum communication. 
In particular, for space-based applications, it has proven to be the most suitable degree of freedom; in ground-to-satellite links, for instance, polarization is inherently immune to the significant Doppler shifts and timing jitter that would otherwise degrade frequency or time-bin encoded signals \cite{Villoresi2008,Yin2017Micius,Yin2020MiciusQKD}.

In this context, fast and actively reconfigurable polarization state measurements are a key requirement for photonic quantum information protocols that rely on real-time, dynamically selected measurement bases. This includes Bell tests with spacelike-separated basis choices \cite{giustina2015significant}, entanglement-based and device-independent (DI) quantum key distribution (QKD) \cite{BBM92,Ekert1991QKD}, feed-forward operations in quantum teleportation \cite{Bennett1993,Ma2012Canaries,Ren2017Micius}, and entanglement swapping \cite{Zukowski1993Swapping,Pan1998FirstSwapping}. In these scenarios, the basis selection rate must match the repetition rate of the source to preserve the statistical independence between subsequent measurement events. 
In loophole-free Bell tests, higher measurement-setting rates reduce experimental latencies, consequently reducing the minimum physical distance required by locality constraints.
In QKD applications, fast reconfigurable receivers enable active switching of the measurement basis, reducing the number of required channels from four to two, since the measurement basis can be actively changed at every trial.

Practical high-speed analysis of polarization states remains technically challenging. Conventional bulk polarization analyzers based on mechanically rotated waveplates cannot be reconfigured at high rates, while electro-optic modulators (EOMs) based on Pockels cells require high driving voltages and exhibit inherent slew rate limitations, thereby limiting their operation to the MHz regime \cite{scheidl2010violation,shalm2015strong,giustina2015significant,bishop2006subnanosecond}. Currently, no solution is available for applications requiring GHz speeds.

Here, we introduce a fiber-based actively reconfigurable polarization state analyzer based on a Sagnac interferometer incorporating a lithium niobate EOM. This system enables active basis selection at rates up to 1 GHz, validated through a CHSH Bell test~\cite{PhysRevLett.23.880} using a 1024-bit pseudorandom sequence. To the best of our knowledge, this represents the highest repetition rate for dynamically reconfigurable polarization-encoded Bell test reported.

The state analyzer is implemented entirely with commercial off-the-shelf (COTS) fiber-optic components and uses phase modulation inside a Sagnac loop to dynamically control the analyzed polarization state. 
In this configuration, the system enables projective polarization measurements on the equator of the Bloch sphere, with the measurement basis determined by the voltage applied to the EOM. This operation is equivalent to that commonly realized with a tunable waveplate.   
Compared with conventional Pockels cell-based analyzers, which typically require driving voltages orders of magnitude larger, the proposed approach substantially simplifies the driving electronics while extending active polarization analysis to the GHz regime.

\section{Theoretical model}
\label{sec: Theoretical model}

\begin{figure}
    \centering
    \includegraphics[width=\linewidth]{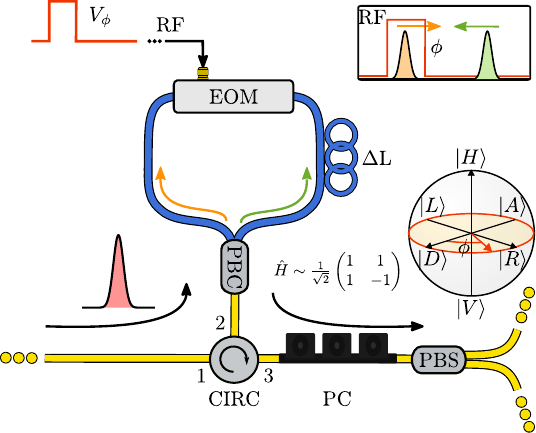}
    \caption{Schematic of the polarization state analyzer. The input pulse enters the circulator (CIRC) through port 1 and proceeds through port 2 into the Sagnac loop for phase modulation. After propagating through the loop, the pulse re-enters port 2 of the CIRC and is routed to port 3, passing through a polarization controller (PC) implementing the Hadamard operation before reaching the final polarizing beamsplitter (PBS). Fiber types are color-coded as blue for polarization maintaining (PM) and yellow for single-mode (SM).}
    \label{fig:Receiver}
   
\end{figure}

The polarization state analyzer, shown in Fig.~\ref{fig:Receiver}, is based on a Sagnac interferometer and uses a modulation scheme similar to the POGNAC \cite{agnesiAllfiber2019}.

A photon in an arbitrary polarization state 
\begin{equation}
    \ket{\psi} = \alpha \ket{H} + \beta \ket{V},
\end{equation}
with $\alpha,\beta \in \mathbb{C}$ and $|\alpha|^2 + |\beta|^2 = 1$, enters the receiver through port 1 of a single-mode (SM) optical circulator (CIRC) and is directed to port 2, where it reaches a polarizing beam combiner (PBC). 
The PBC separates the incoming state into its orthogonal polarization components $\ket{H}$ and $\ket{V}$, both aligned to the slow axis of the polarization-maintaining (PM) fiber forming the Sagnac loop. 
The two components counter-propagate through the loop and encounter an EOM. 
By introducing a suitable fiber delay on one side of the EOM in the Sagnac loop, the two counter-propagating modes reach the EOM at different times, allowing independent phase modulation.

We consider the case in which a relative phase $\phi$ is applied to the $\ket{V}$ component with respect to the counter-propagating $\ket{H}$ component. 
The state before the recombination at the PBC becomes
\begin{equation}
    \ket{\psi'} = \alpha \ket{H} + e^{i\phi} \beta \ket{V}.
\end{equation}
After the recombination, the PBC implements the transformation $\ket{H} \rightarrow -\ket{V}$ and $\ket{V} \rightarrow \ket{H}$, yielding
\begin{equation}
    \ket{\psi''} = \beta \ket{H} - e^{-i\phi} \alpha \ket{V}.
\end{equation}
This overall operation can be described by the unitary transformation $i\sigma_y R_{ps}(\phi)$, where $R_{ps}(\phi)$ denotes the phase shift induced by the EOM \cite{Avesani:20}.

The output state is then directed back to port 2 of the circulator and routed to port 3, where it encounters a polarization controller (PC). The PC is adjusted to transform the polarization basis defined by the PBC into a balanced superposition at the input of the polarizing beamsplitter (PBS). This operation corresponds to a Hadamard transformation ($H$) up to an unknown relative phase $\Delta$, and can be modeled as $R_{ps}(\Delta) H$.

The overall unitary transformation before the PBS is therefore
\begin{equation}
    U_{\text{tot}}(\phi,\Delta) = i \, R_{ps}(\Delta) \, H \, \sigma_y \, R_{ps}(\phi).
\end{equation}

The PBS performs a projective measurement in the $\sigma_z$ basis, with projectors
\begin{equation}
    P^\pm = \frac{\mathbb{I} \pm \sigma_z}{2}.
\end{equation}

\begin{figure*}[ht!]
    \centering
    \includegraphics[width=\linewidth]{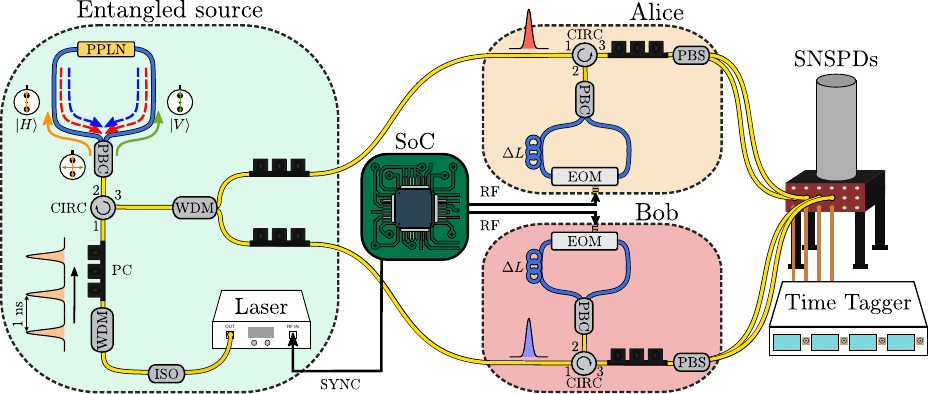}
    \caption{Schematic of the experimental setup. Polarization-entangled photon pairs are generated by a fiber-based non-degenerate source in a Sagnac-loop configuration, pumped by a 1~GHz actively mode-locked laser.
    The non-degenerate photons, depicted in red (signal) and blue (idler), are separated by a wavelength-division multiplexer (WDM) and sent to the Sagnac receivers (Fig.~\ref{fig:Receiver}), labelled Alice and Bob.
    A system-on-chip (SoC) generates and controls the electrical pulses driving the polarization modulation, and the photons are detected using superconducting nanowire single-photon detectors (SNSPDs). ISO: isolator, CIRC: circulator, PBC:  polarizing beam combiner, PBS: polarizing beamsplitter.}
    \label{fig:ExperimentalSetup}
\end{figure*}

The corresponding effective projectors after the state analyzer transformation are given by
\begin{equation}
    P^\pm_\text{eff}(\phi,\Delta) = \frac{\mathbb{I} \pm U_{\text{tot}}^\dagger(\phi,\Delta)\sigma_z U_{\text{tot}}(\phi,\Delta)}{2}.
\end{equation}

Expanding the unitary transformation, we obtain
\begin{equation}
    U_{\text{tot}}^\dagger(\phi,\Delta)\sigma_z U_{\text{tot}}(\phi,\Delta) 
    = U^\dagger(\phi)\, \sigma_z \, U(\phi),
\end{equation}
where $U(\phi) = i\, H\, \sigma_y\, R_{ps}(\phi)$. Since rotations around the $\mathbb{Z}$ axis leave $\sigma_z$ invariant, the effective measurement is independent of the phase shift $R_{ps}(\Delta)$.

Using the Pauli algebra, one finally obtains
\begin{equation}
    U^\dagger(\phi)\sigma_z U(\phi) = -\cos\phi \, \sigma_x + \sin\phi \, \sigma_y.
\end{equation}
Therefore, the measurement implemented by the receiver is equivalent to a projective measurement along the Bloch-sphere axis
\begin{equation}
    \hat{n}(\phi) = (-\cos\phi, \, \sin\phi, \, 0),
\end{equation}
lying in the $\mathbb{X}-\mathbb{Y}$ plane of the Bloch sphere.

In particular, for $\phi = 0$ and $\phi = \pi/2$, the measurement corresponds to projections onto the eigenstates of $\sigma_x$ and $\sigma_y$, respectively.

For a CHSH Bell test, Alice can choose measurement settings $\phi^A \in \{0, \pi/2\}$, corresponding to the observables $\{-\sigma_x, \sigma_y\}$, while Bob can choose $\phi^B \in \{-\pi/4, \pi/4\}$, corresponding to the observables
\begin{equation}
    \left\{- \frac{\sigma_x + \sigma_y}{\sqrt{2}}, \frac{\sigma_y - \sigma_x}{\sqrt{2}} \right\}.
\end{equation}

\begin{figure*}[ht!]
    \centering
    \includegraphics[width=\linewidth]{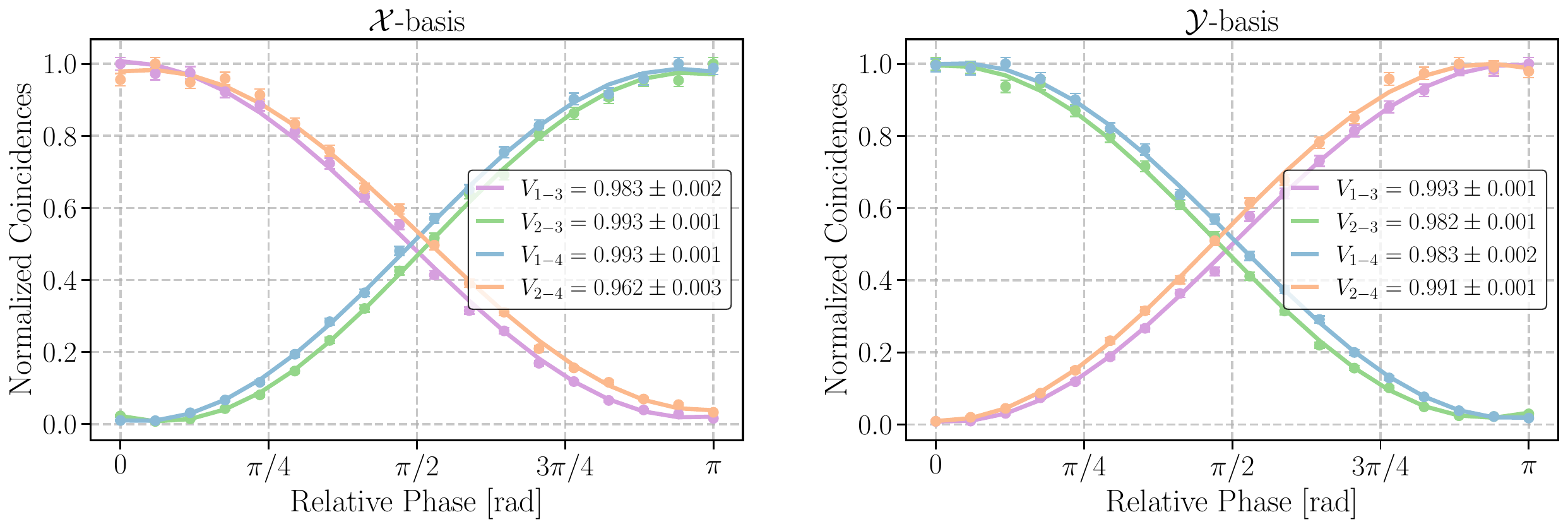}
    \caption{Visibility scans of the entangled states in the two mutual unbiased bases $\mathcal{X}$ and $\mathcal{Y}$. These scans are obtained by modulating Alice at 0 (for $\mathcal{X}$-basis) and $-\pi/2$ (for $\mathcal{Y}$-basis), while Bob changes the relative phase between the receivers from 0 to $\pi$.   }
    \label{fig:Visibilities_X_Y}
\end{figure*}

\section{Experimental Setup}
\label{sec: Experimental Setup}

To experimentally validate the performance of the proposed receiver in high-speed regime, we designed and implemented a laboratory experiment based on a CHSH Bell test operating at 1 GHz. The experimental setup comprises a fiber-based polarization entangled photon source and two Sagnac-based receivers, with the entire architecture implemented using COTS fiber-optic components.
The experimental setup is shown in Fig. \ref{fig:ExperimentalSetup}.
We employ a fiber-based, non-degenerate polarization-entangled photon source in a Sagnac-loop configuration.
Photon pairs at telecom wavelengths are generated inside the Sagnac interferometer via a cascaded SHG+SPDC process \cite{Arahira:11, Coccia:2025vik} in a type-0 periodically poled lithium niobate (PPLN) waveguide. 
Pump photons are centered at $\lambda_p = 1560.61$~nm (ITU channel 21) while the generated photon pairs (signal and idler) are cenetered at wavelengths $\lambda_s = 1558.17$~nm (ITU channel 24) and $\lambda_i = 1563.05$~nm (ITU channel 18). 
The cascaded process can be described as
\begin{equation}
    \ket{V}_{\lambda_p}\ket{V}_{\lambda_p} \xrightarrow{\text{SHG}} \ket{V}_{\lambda_{\text{SHG}}} \xrightarrow{\text{SPDC}} \ket{V}_{\lambda_s}\ket{V}_{\lambda_i}.
\end{equation}
The source is pumped by an active mode-locked laser, producing picosecond-duration pulses at a repetition rate of 1~GHz. 
The linearly polarized pump light is coupled into an SM fiber.
It is spectrally filtered using 100~GHz bandwidth wavelength-division multiplexing (WDM) filters to suppress out-of-band noise.
The light is then directed towards the Sagnac loop.
This is achieved by injecting it into port 1 of a CIRC and routing it to port 2, after passing through a PC (see Fig.~\ref{fig:ExperimentalSetup}).

The PC is adjusted to generate a balanced superposition of horizontally and vertically polarized light at the input of the PBC of the Sagnac loop.
The PBC equally splits the pump into the two counter-propagating directions of the Sagnac loop, where the light propagates along the slow axis of PM fibers. 
The generation axis of the nonlinear crystal is aligned to the slow axis, enabling efficient cascaded SHG+SPDC.

Photon pairs generated in the clockwise and counterclockwise directions recombine at the PBC, leading to polarization components $\ket{H}_{\lambda_s}\ket{H}_{\lambda_i}$ and $\ket{V}_{\lambda_s}\ket{V}_{\lambda_i}$, depending on the recombination port. The resulting two-photon state is
\begin{equation}
\label{eq: entangled states}
    \ket{\psi} = \frac{1}{\sqrt{2}}\left(\ket{H}_{\lambda_s}\ket{H}_{\lambda_i} + e^{i\varphi}\ket{V}_{\lambda_s}\ket{V}_{\lambda_i} \right),
\end{equation}
with a relative phase $\varphi$ that depends on the PC setting.

The generated photons exit the PBC and are directed back to port 2 of the circulator, which routes them to port 3. 
Here, spectral filtering is performed to remove the residual pump, while signal and idler photons are separated into two distinct paths.

The signal and idler photons are directed to two independent Sagnac-based polarization receivers (see Fig.~\ref{fig:Receiver}).
Each receiver incorporates a thin-film lithium niobate EOM with a bandwidth of 20~GHz. 
The modulation is driven by electrical pulses with a duration of 166.66~ps, generated by a system-on-chip (SoC) equipped with a 6~GSa/s digital-to-analog converter (DAC). 
The electrical output signals of the SoC are then amplified using RF amplifiers. 
In the current implementation, the SoC generates the voltage waveforms for phase-modulation scans during visibility measurements. 
However, the CHSH Bell test measurements require only two modulation settings, allowing for simpler and more efficient electronic control architectures.

A path-length asymmetry of $\Delta L \approx 10$~cm is introduced inside each Sagnac loop, corresponding to a temporal delay of $\Delta t \approx 500$~ps between the counter-propagating modes.
The maximum achievable modulation rate is mainly determined by the Sagnac-loop asymmetry, the electrical pulse width, and the EOM length. In our case, following the model proposed by Berra \textit{et al.} \cite{berra2025generalmodelmodulationstrategies},
and considering an EOM propagation time of $\tau_{\mathrm{EOM}} \approx 250$~ps, we estimate a maximum modulation rate of $\approx1.2$~GHz.
An alternative approach to further extend the operational frequency beyond the limits of the architecture considered in this work is given by the symmetric configuration proposed in \cite{berra2025generalmodelmodulationstrategies}.

Polarization controllers at the input of each receiver are used to compensate for fiber-induced transformations and to adjust the relative phase of the entangled state, ensuring that the input state is the Bell state $\ket{\Phi^+} = \frac{1}{\sqrt{2}}(\ket{HH} + \ket{VV})$.
The outputs of each receiver are connected to superconducting nanowire single-photon detectors (SNSPDs). 
The photon arrival times are recorded using a time tagger with 1~ps resolution and 6~ps RMS jitter.

\begin{table*} [ht!]
\centering
\footnotesize
\renewcommand{\arraystretch}{1.15}

\begin{tabular*}{\textwidth}{@{\extracolsep{\fill}} l c c c c}
\toprule
\toprule
\textbf{Preceding symbol} & $\Delta \langle A_0 B_0 \rangle$ & $\Delta \langle A_0 B_1 \rangle$ & $\Delta \langle A_1 B_0 \rangle$ & $\Delta \langle A_1 B_1 \rangle$ \\
\midrule

$A_0 B_0$ & $-0.007 \pm 0.004$ & $0 \pm 0.004$ & $0.009 \pm 0.004$ & $0.002 \pm 0.003$ \\

$A_0 B_1$ & $-0.004 \pm 0.004$ & $0.010 \pm 0.004$  & $-0.003 \pm 0.004$ & $ 0 \pm 0.003$ \\

$A_1 B_0$ & $0.002 \pm 0.004$ & $-0.007 \pm 0.004$  & $0.001 \pm 0.004$ & $0.001 \pm 0.003$ \\

$A_1 B_1$ & $0.010 \pm 0.003$ & $-0.003 \pm 0.004$  & $-0.007 \pm 0.004$ & $-0.003 \pm 0.003$ \\

\bottomrule
\bottomrule
\end{tabular*}

\caption{Deviation of each correlator value $\Delta \langle A_i B_j  \rangle$ from its mean value, conditioned on the preceding symbol. The observed deviations are compatible to zero and can be therefore mainly attributed to statistical fluctuations.}
\label{tab: deviation_from_mean_correlators}
\end{table*}

\section{Results}
\label{sec:Results}

We first characterized the receivers by performing the two-photon visibility scan of the polarization-entangled state $\ket{\Phi^+}$.
The scan was performed by varying the measurement basis of one receiver (Alice) by adjusting the relative phase $\phi$ from $0$ to $\pi$, while keeping the second receiver (Bob) fixed.
For the $\mathcal{X}$-basis, Alice scans $\phi_A$ in $[0,\pi]$ and Bob is set to $\phi_{B} = 0$. For the $\mathcal{Y}$-basis basis Alice scans $\phi_A$ in $[-\pi/2,\pi/2]$ and Bob is set to $\phi_{B} = -\pi/2$.
The results demonstrate high visibilities of up to $99\%$ in both the $\mathcal{X}$ and $\mathcal{Y}$ bases (Fig.~\ref{fig:Visibilities_X_Y}).

\begin{figure}[b!]
    \centering
    \includegraphics[width=\linewidth]{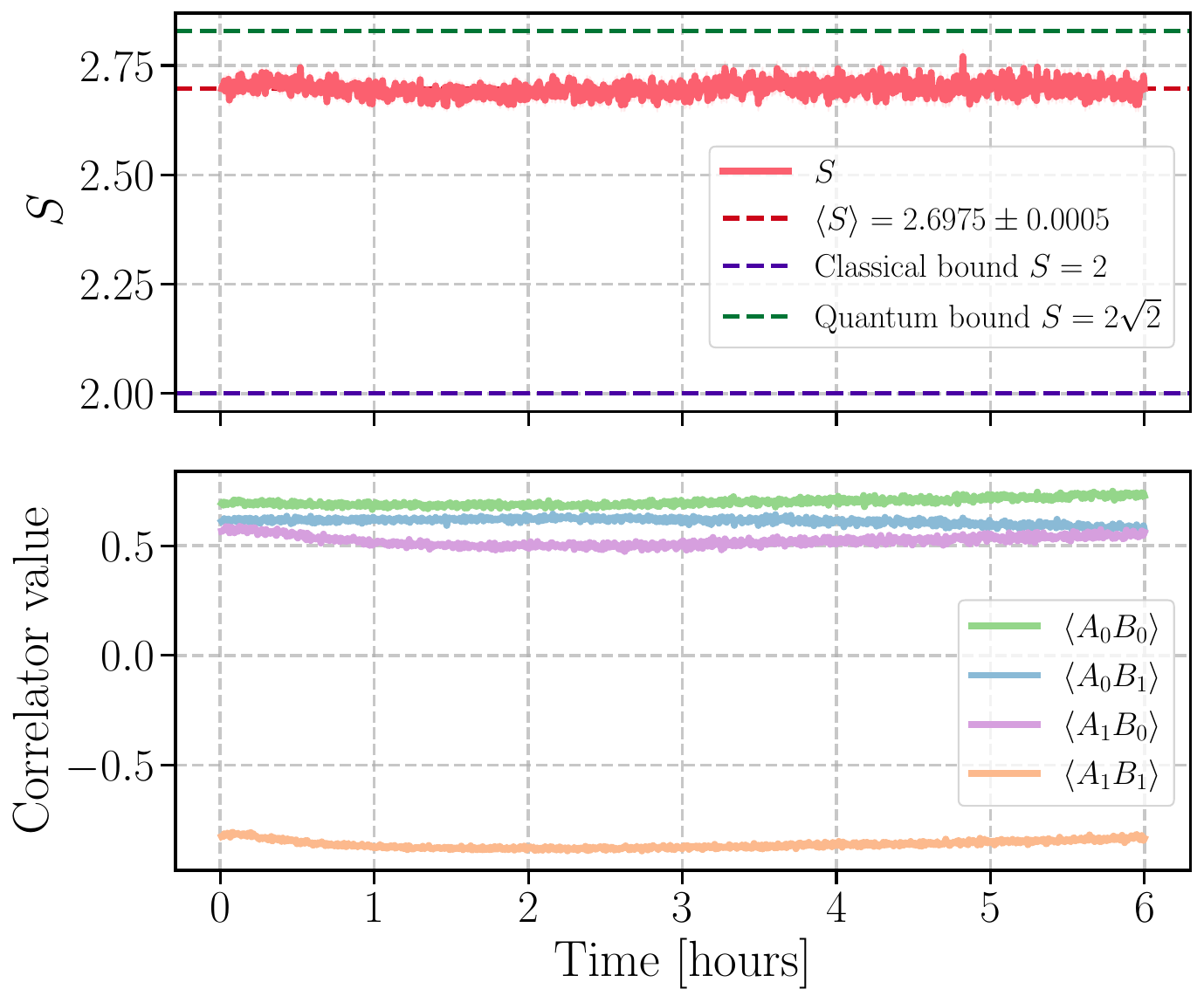}
    \caption{Long-term stability of the system during a 1 GHz CHSH test driven by a 1024-bit pseudo-random sequence. The plots show the evolution over time of the S parameter (top panel) and the CHSH correlators (bottom panel).}
    \label{fig:CHSH_correlators_vs_time}
\end{figure}

To validate the high-speed performance of the receiver, we performed a CHSH Bell test in which the measurement settings were varied on a trial-by-trial basis at a repetition rate of 1~GHz, following a 1024-bit pseudo-random sequence.
Alice’s input phase was chosen from the set $\{0, \pi/2\}$, while Bob’s was chosen from $\{-\pi/4, \pi/4\}$.
To assess the long-term stability of the system, the CHSH measurement was continuously performed over a duration of 6~hours.
Throughout this measurement, the system maintained a stable Bell violation, as shown in Fig.~\ref{fig:CHSH_correlators_vs_time}, demonstrating its robustness for practical quantum communication applications.
An average Bell parameter of $S = 2.6975 \pm 0.0005$ was obtained over the whole 6~hours measurement.

The receiver presented here offers a significant advantage over conventional bulk Pockels cell-based implementations, which are typically limited to the MHz regime due to the high driving voltages required, on the order of hundreds of volts \cite{scheidl2010violation,shalm2015strong,giustina2015significant,bishop2006subnanosecond, Ma2012Canaries}.
These constraints impose significant electronic and physical limitations, ultimately restricting the achievable repetition rates. On the electronics side, the driving circuitry must sustain high-voltage high-speed transients, leading to stringent requirements in terms of RF power handling, impedance matching, and thermal dissipation.
On the other hand, bulk electro-optic crystals have an intrinsic damage threshold, which depends on the specific material and limits the sustainable operation frequency to the low-megahertz range.
In contrast, in the traveling wave lithium niobate waveguide EOMs as used here, the voltage required to achieve a $\pi$ phase shift is on the order of a few volts; well within reach of FPGA-generated signals amplified by COTS RF amplifiers. 

In the present implementation, accurate modulation requires precise synchronization between the electrical driving signals and the optical pulses, as any timing mismatch can induce unintended phase shifts, affecting the measurement outcomes and reducing the observed CHSH violation.
At sufficiently high repetition rates, active polarization analyzers may also exhibit patterning effects, i.e., correlations between consecutive measurements arising from residual inter-symbol interference \cite{cryst9010049}. 
To assess the possible presence of these effects, in Tab.~\ref{tab: deviation_from_mean_correlators} we report the deviation of the correlator values for each measurement setting $(i,j)$ of Alice and Bob, conditioned on the preceding settings $(k,m)$, namely $\langle A_i B_j \mid A_k B_m \rangle$ with respect to the corresponding mean value. 
At~1 GHz, the measured deviations are consistent with statistical uncertainties indicating no observable patterning effects. 
A more comprehensive analysis of these limitations at higher repetition rates, together with mitigation strategies for Sagnac-based modulators, is presented in~\cite{berra2025generalmodelmodulationstrategies}.

\section{Conclusions}
\label{sec:Conclusions}

In this work, we demonstrated a novel, fully fiber-based high-speed active polarization state analyzer. The system was experimentally validated as a receiver for a Bell test with trial-by-trial reconfiguration of the measurement basis. The proposed receiver is based on lithium niobate EOMs, which operate at bandwidths on the order of tens of GHz, in contrast with bulk Pockels cells commonly employed in entanglement certification experiments with active basis choice, typically limited to repetition rates of a few MHz.
By varying the applied phase, the receiver enables measurements over different bases on the equator of the Bloch sphere. Using this receiver, we achieved two-photon visibilities of up to $99\%$ and performed a Bell test at a repetition rate of 1~GHz which is, to the best of our knowledge, a record in terms of measurement speed for polarization entangled photons.
The system exhibits high robustness, as demonstrated by a long-term stability measurement over 6~hours with a Bell parameter of $S = 2.6975 \pm 0.0005$.

Beyond entanglement certification, the proposed receiver represents a versatile solution for a broad range of quantum communication protocols requiring fast and reconfigurable polarization measurements. This includes QKD where such a device could find several applications. Firstly, it could be used for active basis selection at the receiver, reducing the number of required detectors without employing temporal multiplexing schemes that limit the maximum achievable repetition rate. Additionally, active basis selection could allow for optimization of the basis choice probability as a function of channel and hardware conditions~\cite{Rusca2018}. Furthermore, this receiver scheme can be used as a counter-measurement for the detector efficiency quantum hacking attack vector \cite{PhysRevA.78.042333} since this vulnerability can be mitigated by scrambling the physical detector allocation, i.e., changing for each trial whether a click corresponds to a 0 or 1 outcome.
The receiver could also be employed in event ready physics experiments such as the instrumental test \cite{Chaves2018} or for photonic quantum state teleportation \cite{Bouwmeester1997} since in both cases a unitary transformation that depends on the outcome of a measurement has to be applied to the quantum states. 
Lastly, the presented device could potentially be exploited for DI protocols. For such applications, a key practical limitation is the insertion losses of the commercially available lithium niobate EOMs, which are typically in the order of 2--4~dB. Insertion losses are primarily dependent on the fiber-to-chip coupling of the phase modulators due to their significant mode-field mismatch. Recent advances to optimize these coupling losses down to sub-decibel levels include using tapered single-mode fibers with optimized modal profiles \cite{Yao:20}, integrating inverse taper structures and lithography-friendly spot-size converters (SSCs) on-chip to expand the tightly confined mode field \cite{Shen:26}, or employing double-side irradiation self-written waveguides \cite{He:23}. By transitioning to suitable low-loss \cite{Pereira2022Electrooptic} and all-optical switching architectures \cite{Kupchak2019Terahertz, Couture2025Terahertz}, the losses could be minimized, rendering the presented scheme compatible with the requirements for high-speed DI protocols.

\section*{Fundings}
This project has received funding from the European Union’s Horizon Europe research and innovation programme under the project ``Quantum Secure Networks Partnership'' (QSNP, grant agreement No 101114043).
 Views and opinions expressed are however those of the author(s) only and do not necessarily reflect those of the European Union or European Commission-EU. Neither the European Union nor the granting authority can be held responsible for them.

\section*{Acknowledgements}
The authors would like to acknowledge Matías Rubén Bolaños for developing the SoC control software and Andrea Curiale for assistance with the experimental setup.
K.V. acknowledges support from the John Templeton Foundation by grant No. 63132 as a recipient of the Enrico Fermi Fellowship awarded through the Center for SpaceTime and Quantum.

\section*{Author contributions}

All authors designed and implemented the system and the experiment.
A.P and K.V performed the measurements and analyzed the data.
All authors contributed to the writing and review of the manuscript.

\bibliography{biblio}

\onecolumngrid
\appendix

\end{document}